\documentclass[9pt,twocolumn,twoside]{osajnl}

\journal{pr} 

\setboolean{shortarticle}{false}

\title{Ultra-short pulses with high repetition frequency in transmission plasmonic systems}

\author[1]{Shilei Li}
\author[1,*]{Fei Xing}
\author[2,*]{Li Yu}

\affil[1]{School of Physics and Optoelectronic Engineering, Shandong University of Technology, Zibo 255049, China}
\affil[2]{State Key Laboratory of Information Photonics and Optical Communications, School of Science, Beijing University of Posts and Telecommunications, Beijing 100876, China}

\affil[*]{Corresponding author: yuliyuli@bupt.edu.cn}

\begin{abstract}
Ultra-short pulses with high repetition frequency have great application prospects in the field of nano-optics. Here, in the case of continuous wave incidence, the femtosecond pulses with THz repetition frequency are achieved in the transmission system consisting of a rectangular cavity, a V-groove (VG) cavity and a nanowire embedded with quantum emitters (QEs). The generation mechanism of the ultra-short pulses with high repetition frequency is elucidated by semi-classical Dicke model. Attribute to the presence of the two-level QEs, the field amplitude in plasmonic resonator is oscillating with time, resulting in the transmittance of the system behave as the form of pulse oscillation. Moreover, The pulse repetition frequency and extinction ratio can be freely controlled by the incident light intensity and QEs number density to obtain the required ultra-short pulses at nanoscale, which also has potential applications in optical computing.
\end{abstract}

\setboolean{displaycopyright}{true}

\begin{document}

\maketitle

\section{Introduction}
Research on nanolasers has made great progress since Bergman and Stockman proposed the concept of spaser (Surface Plasmon Amplification by Stimulated Emission of Radiation) in 2003 \cite{r1,r2,r3,r4}. However, achieving ultra-short pulses with high repetition frequency is still one of the main topics to be investigated and resolved urgently in the field of nano-integrated optics. Photonic crystal (PC) nanolasers operating under electrical injection inevitably introduced both scattering and absorption losses, which also has complications of poor heat conduction and low mechanical stability \cite{r5,r6}. The miniaturization of semiconductor nanowire lasers is limited by the diffraction limit, so it is extremely difficult to further reduce the resonator size \cite{r7}. Many studies show that metal-based nanostructures become the key ingredient to solve this problem \cite{r8,r9,r10,r11}. Because of they have ability to break the diffraction limit \cite{r12}, and the light can also be localized to the nanoscale, thereby enhancing the interaction between light and quantum emitters (QEs) \cite{r13,r14}. And the investigations on the coupling between metal-based nanostructures and QEs also indicate that the hybrid system of plasmonic resonator and QEs is very competitive for obtaining ultra-short pulses with high repetition frequency at the nanoscale.

Meanwhile, the interaction between light and QEs is studied deeply and comprehensively in recent years \cite{r15,r16,r17,r18}, and the strong coupling between metallic nanostructures and QEs has been experimentally realized \cite{r19,r20}. The ultrafast Rabi oscillations between excitons and plasmons were also observed in metal nanostructures with J-aggregates \cite{r21}. Whereafter, A. Demetriadou et al. investigated the spatiotemporal dynamics and control of strong coupling in plasmonic nanocavities \cite{r22}. R. Liu and J. Ren et al. studied the strong interaction between light and QEs in open plasmonic nanocavity systems \cite{r23,r24}, and very significant Rabi splitting were observed at room temperature. These investigations laid the foundation for the interaction between light and QEs to be applied in nano-integrated optics. And we find that the coupling between plasmonic resonators and QEs can also be used to obtain femtosecond pulses with high repetition frequency in the case of continuous wave incidence, which is rarely reported in plasmonic resonator systems. Moreover, traditional methods are unable to satisfy the application requirements of many high-speed signal processing fields at the nanoscale, such as optical communications \cite{r25}, optical interconnection \cite{r26,r27,r28,r29}, and on-chip sensing \cite{r30,r31,r32,r33}. So, femtosecond pulses with high repetition frequency are highly desirable at nanoscale.

\begin{figure*}[hbtp]
\centering\includegraphics[width=14cm]{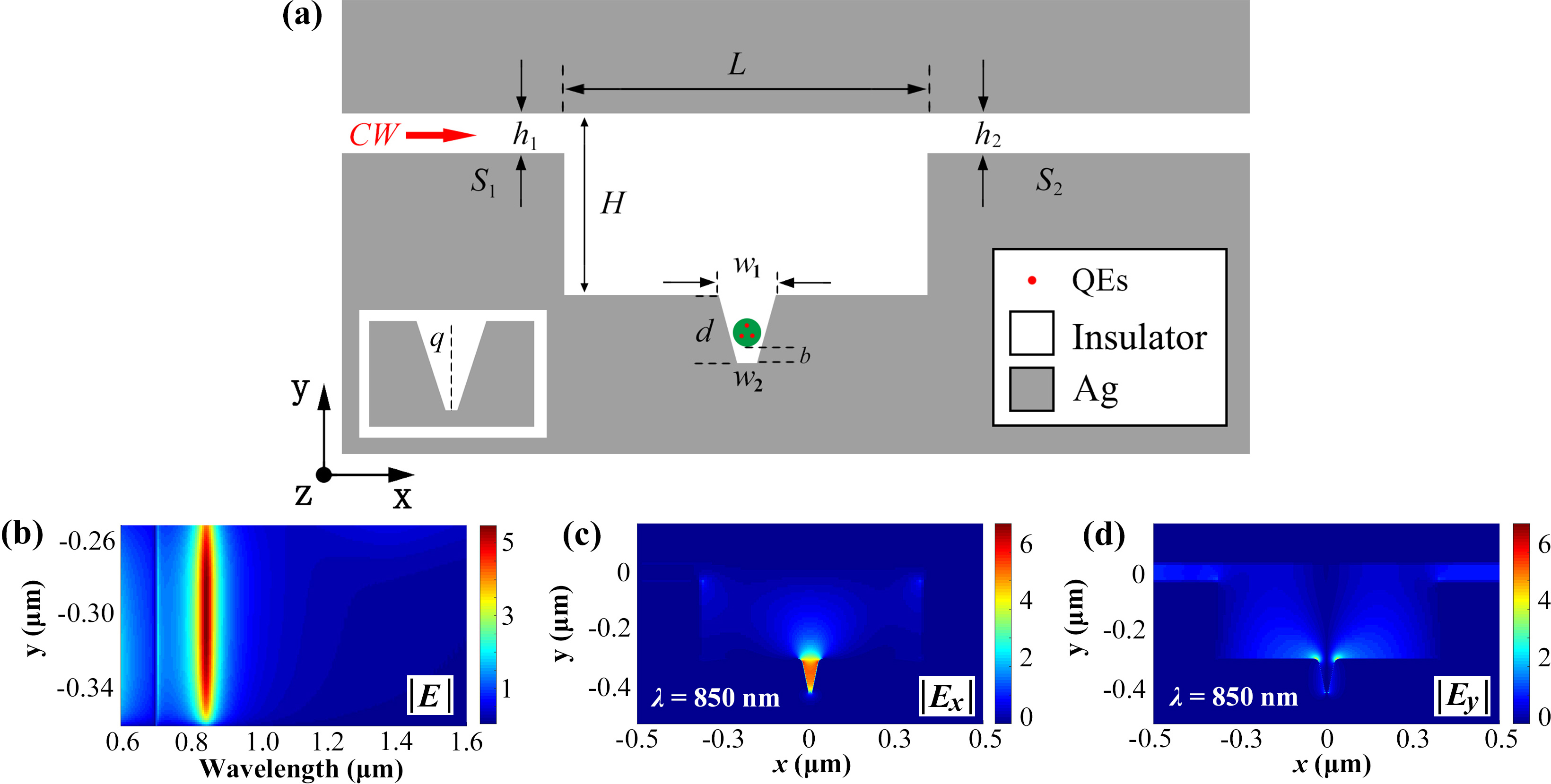}
\caption{(a) Schematic diagram of the hybrid system of plasmonic resonator and QEs. The insulator is set to air with a refractive index of 1. The metal around the VG cavity is silver, the dielectric function of which is obtained from the experimental data of Palik \cite{r34}. The widths of waveguides $S_1$ and $S_2$ are $h_1$ = 50 nm and $h_2$ = 50 nm. The length and height of the rectangular cavity are $L$ = 640 nm and $H$ = 280 nm, respectively. The width and depth of the VG cavity are $w_1$ = 50 nm and $d$ = 102 nm, the bottom width is $w_2$ = 10 nm. The diameter of the nanowire is 16 nm, and the distance from its bottom to the bottom of the VG cavity is $b$ = 34 nm. (b) Distribution of electric field $|E|$ on center line \textit{q} (in the left inset of Fig. 1(a)) at different wavelengths. (c) and (d) Distribution of the electric field components $|E_x|$ and $|E_y|$ at the resonant wavelength of 850nm.}
\label{Fig_1}
\end{figure*}

In this paper, we numerically and theoretically studied the transmission response of a hybrid system composed of plasmonic resonators and a nanowire embedded with two-level QEs. Under the weak excitation limit, obvious mode splitting appears in the transmission spectrum of the system, which indicates that the coupling between the plasmonic resonators and QEs is prominent. However, when the intensity of the excitation light is relatively strong, the QEs in nanowire will oscillate between the upper and lower energy levels, and the collective population difference function $Sz(t)$ cannot be approximated as a constant again, but a function of time. This means the output power of the nanowire is in the form of oscillation, which resulting in the transmittance of the system also oscillate with time in the form of pulse. Moreover, the pulse repetition frequency and extinction ratio can be freely controlled by the incident light intensity and the QEs number density to achieve the required pulse form. This provides a method to obtain ultra-short pulses with high repetition frequency at the nanoscale.

\section{Structure and Field Distribution}

The investigated hybrid system of plasmonic resonator and QEs is shown in Figure 1, which consists of two waveguides, a rectangular cavity, a V-groove (VG) cavity and a nanowire embedded with two-level QEs. The complex dielectric function of the QEs embedded on the nanowire is described by the Lorentz model \cite{r35}: $\varepsilon(\omega)=\varepsilon_{\infty}-f\omega_{L}^2/(\omega^2-\omega_{L}^2+i\gamma\omega)$. The role of the nanowire is to make the distribution of the QEs more concentrated. Without the nanowire, the distribution of QEs is difficult to concentrate. The incident wave is continuous wave (CW) in transverse magnetic (TM) mode, which is input from the waveguide S$_1$ on the left. The aim of our investigation is the influences of the incident light intensity and the QEs number density on the transmission response of the hybrid system in the case of continuous wave incidence together with the evolution of the transmittance over time.

In order to obtain stronger coupling between the plasmonic resonator and QEs, the electric field distribution without nanowire in VG cavity is investigated first. Fig. 1(b) shows the pseudo-color image of the electric field distribution on the centerline \textit{q} of the VG cavity in Fig. 1(a) utilizing two-dimensional FDTD method, the incident light intensity is set to 1. When the wavelength of the incident light is 850 nm, the electric field on the center line \textit{q} of the VG cavity reaches maximum value at the position about 60 nm from the bottom of the VG cavity. Figs. 1(c) and 1(d) respectively show the distribution of the electric field components $|E_x|$ and $|E_y|$ in the plasmonic resonator when the incident light wavelength is 850nm. The electric field energy in plasmonic resonator is mainly concentrated in the VG cavity, the energy in rectangular cavity is close to zero, and the electric field component in VG cavity is mainly $|E_x|$, component $|E_y|$ is almost zero. Therefore, in order to obtain stronger coupling between plasmonic resonator and QEs, the transition dipole moment of the QEs should be as parallel as possible to the x-axis.

\section{Weak excitation limit}

Before studying the transmission response of the hybrid system consisting of plasmonic resonator and QEs, it is necessary to investigate the transmission response of the system without nanowire in VG cavity in order to obtain the coupling parameters between the waveguide and the resonant mode in plasmonic resonator. These coupling parameters are necessary for analyzing the evolution of the transmittance with time in Section 4. When the two waveguides are symmetrical about the rectangular cavity and equal in width, according to the multimode interference coupled mode theory (MICMT) \cite{r36}, the transmission coefficient of the coupled system can be expressed as follows

\begin{equation}  
t = \sum_{m} \frac{2e^{i\varphi_m}}{-i(\omega-\omega_m)\tau_m+2+\frac{\tau_m}{\tau_{m0}}}
\label{Eq_1}
\end{equation}
Then, the transmittance of the coupled system is $T = |t|^2$. Where, $\omega_m$ is the resonant angular frequency of the m-th mode of the plasmonic resonator; $\tau_m$ is the coupling decay time between the waveguide and the m-th resonant mode, $\tau_{m0}$ is the decay time of the internal loss. $\varphi_m$ is the total phase difference between the waveguide and the m-th resonant mode.

\begin{figure}[hbtp]
\centering\includegraphics[width=8.6cm]{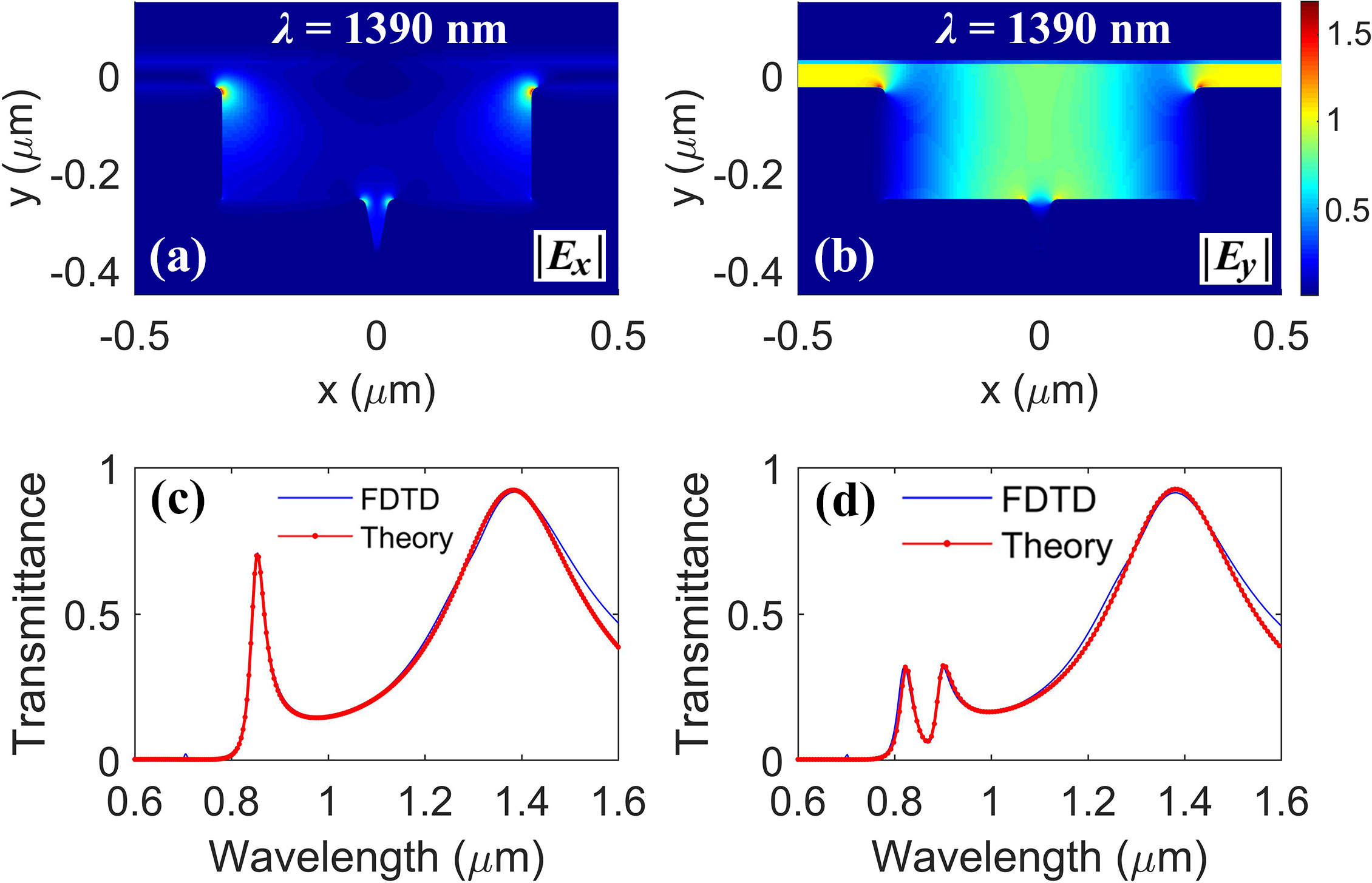}
\caption{(a) and (b) The distribution of the electric field components $|E_x|$ and $|E_y|$ at the resonant wavelength of 1390 nm. (c) The simulation (blue) and theoretical (red) curves of the transmittance of the coupled system without nanowire in VG cavity. The fitting parameters are $\tau_1$ = 56 fs,\; $\tau_2$ = 13 fs,\; $\tau_{10}$ = 100 fs,\; $\tau_{20}$ = 120 fs,\; $\varphi_1$ = 0.4$\pi$,\; $\varphi_2$ = -0.5$\pi$, respectively. (d) The simulation (blue) and theoretical (red) curves of the transmittance of the hybrid system with nanowire in VG cavity. Here, the Lorentz model parameters of the QEs are $\varepsilon_\infty$ = 1.4,\; $f$ = 0.4,\; $\omega_L=2.10\times10^{15}$ rad/s,\; $\gamma=3.85\times10^{13}$ rad/s, respectively. The coupling strength is $g_1=9\times10^{13}$ rad/s. Since the nanowire is placed in VG cavity, the coupling decay time of the TMv mode slightly changes to $\tau_1$ = 50 fs, while other fitting parameters remain unchanged.}
\label{Fig_2}
\end{figure}

It is well known that there are many resonant modes in plasmonic resonator. In order to facilitate theoretical analysis, the transmission coefficient of each mode is set to be $t_m = \sum_{m} 2e^{i\varphi_m}/{[-i(\omega-\omega_m)\tau_m+2+\tau_m/\tau_{m0}}]$. Here, we only consider two resonant modes that have great impact on the transmission coefficient of the system. One of them is the resonant mode in the VG cavity (the distribution of the electric field component has been shown in Figs. 1(c) and 1(d)), which could be called TMv mode (the amplitude is represented by $a_1$) and the resonant wavelength is 850nm. Another resonant mode is the TM$_{1,0}$ mode (the amplitude is represented by $a_2$) in the rectangular cavity, of which the resonant wavelength is 1390nm and the field distributions of $|E_x|$ and $|E_y|$ are given by Figs. 2(a) and 2(b). It can be seen from Figs. 2(a) and 2(b) that the electric field component of the TM$_{1,0}$ mode in rectangular cavity is mainly $|E_y|$, and the component $|E_x|$ is almost zero. In the range of 600 nm $\sim$ 1600 nm, the transmission response of the system is mainly affected by these two resonant modes. So, the transmittance of the system can be simplified as $T = |t_1+t_2|^2$. The transmittance of the coupled system without nanowire in VG cavity is calculated utilizing 2D FDTD method. Fig. 2(c) shows the simulation (blue line) and theoretical (red line) results of the transmittance, in which there are two peaks on the transmittance curve. The position of the first peak is at 854 nm which can be adjusted by the depth of the VG cavity \cite{r37}, but the width has almost no effect on it. The position of the second one is at 1386 nm and can be adjusted by the length of the rectangular cavity, the height has almost no effect on the position of the second peak \cite{r37}. If there is no rectangular cavity, the mode volume of the VG cavity will be larger, which is not conducive to the coupling between the QEs and VG cavity. The role of the rectangular cavity is to reduce the mode volume of the VG cavity, but changing $H$ and $L$ of the rectangular cavity will not affect the mode volume of the VG cavity, so the coupling between QEs and VG cavity will not be influenced by them.

The parameters of the coupling system without nanowire in VG cavity can be obtained by fitting the theoretical curve to the simulation curve. But for the hybrid system with the nanowire placed in VG cavity, the coupling parameters are slightly different. Based on the research above, we need to further obtain the coupling parameters of this hybrid system for the investigation in section 4. So, next investigated will be the transmission response of the hybrid system under the weak excitation limit when there is a nanowire embedded with QEs placed in the VG cavity.

The QEs system is described by Dicke model, according to the Heisenberg operator equation \cite{r38} and MICMT \cite{r36}, the following coupling equations can be obtained
\begin{align} 
\begin{split}
\frac{da_m}{dt}=&-(i\omega_m+\frac{1}{\tau_{m0}}+\frac{1}{\tau_{m1}}+\frac{1}{\tau_{m2}})a_m \\&-ig_{m,0} S_-+\kappa_{m1}b_{m,1+}+\kappa_{m2}b_{m,2+}
\end{split} \\ 
  \frac{dS_-}{dt}=&-(i\omega_A+\gamma)S_-+iS_z(t)\sum_m{g_{m,0}a_m} \\ 
  b_{1-}=-b_{1+}&+\sum_m{\kappa_{m1}^*a_m},\quad b_{2-}=-b_{2+}+\sum_m{\kappa_{m2}^*a_m}
\end{align}
Where, $b_{1+}=\sum_m{b_{m,1+}}$ and $b_{2+}=\sum_m{b_{m,2+}}$. $\kappa_{m1}=e^{i\theta_{m1}}\sqrt{2/\tau_{m1}}$ and $\kappa_{m2}=e^{i\theta_{m2}}\sqrt{2/\tau_{m2}}$. $\tau_{m1}$ and $\tau_{m2}$ are the decay time of the coupling between the m-th resonant mode and waveguides ($S_1$ and $S_2$), $\theta_{m1}$ and $\theta_{m2}$ are the corresponding coupling phases. $g_{m,0}$ is the coupling strength between a single QE and the m-th mode in plasmonic resonantor. $S_-(t)=\sum_{j=1}^NS_-^{(j)}(t)$ and $S_z(t)=\sum_{j=1}^NS_z^{(j)}(t)$ are the collective dipole operator and population difference operator, respectively. $\omega_A$ is the resonant angular frequency of the QEs, $\gamma$ is the decay rate. $b_{i\pm}=\sum_mb_{m,i\pm}$ are the field amplitudes in each waveguide (\textit{i} = 1,2, for incoming (+) or outgoing (-) from the resonator). In the weak excitation limit, QEs are basically in the ground state \cite{r38}, which means $S_z^{(j)}(t)=-1 \rightarrow S_z(t)=-N$, $N$ is the number of QEs. From Fig. 2(c), it can be seen that the transmission spectrum of the TM$_{1,0}$ ($a_2$) mode is relatively broad. Only from the perspective of the transmission spectrum, the coupling between TM$_{1,0}$ mode and QEs cannot be ignored. However, the coupling strength between QEs and the resonant mode in plasmonic resonator is also closely related to the field distribution. Figs. 2(a) and 2(b) show the distribution of the electric field components $|E_x|$ and $|E_y|$ at the resonant wavelength of 1390 nm, from which it can be found that the electric field energy of the TM$_{1,0}$ mode is mainly concentrated in the rectangular cavity , the energy in VG cavity is almost zero. Since the QEs is placed in VG cavity, the coupling between QEs and TM$_{1,0}$ mode is approximately zero. We can reasonably ignore the coupling between the QEs and TM$_{1,0}$ mode, and only consider the coupling between QEs and the TMv mode of which the resonant frequency is very close to that of the QEs. We assume $g_1=\sqrt{N}g_{1,0}$, for symmetrical system $\tau_{m1}=\tau_{m2}=\tau_m$, the transmission coefficient expression of the hybrid system can be expressed as follows

\begin{eqnarray} 
\begin{split}
t=&\frac{2e^{i\varphi_1}(-i\Delta_A+\gamma)}
{(-i\Delta_1\tau_1+2+\frac{\tau_1}{\tau_{10}})(-i\Delta_A+\gamma)+\tau_1 g_1^2} \\ 
&+{\frac{2e^{i\varphi_2}}
{-i\Delta_2 \tau_2+2+\frac{\tau_2}{\tau_{20}}}}
\end{split}
\label{Eq_5}
\end{eqnarray}
Here $\Delta_{A,1,2}=\omega-\omega_{A,1,2}$. Some parameters in expression (5) (such as $\tau_1$, $\tau_{10}$, $\varphi_1$, etc.) are difficult to be directly given by theory, but need to be provided by the curve-fitting of theory to simulation. Fig. 2(d) shows the simulation (blue) and theoretical (red) curves of the transmittance of the hybrid system with the nanowire in VG cavity embedded with QEs. Obvious mode splitting appears in the transmission spectrum, and the distance (80 nm) between the two split peaks in Fig. 2(d) is much larger than the full width at half maximum (FWHM) (42 nm) of the transmission window of TMv mode in Fig. 2(c), which indicates that the coupling between QEs and the TMv mode of VG cavity is strong coupling. By curve-fitting, the coupling and internal loss decay time of the TMv mode can be obtained as $\tau_1$ = 50 fs and $\tau_{10}$ = 100 fs, which will be used in the research of the non-weak excitation limit in next section.

\section{Non weak excitation limit}

The above analysis and study is performed under the weak excitation limit the case of which the incident light intensity is very weak. As the incident light intensity is stronger, the QEs in the nanowire no longer remain in the ground state. This means the collective population difference operator $Sz(t)$ cannot be approximated as a constant -\textit{N}, but a function changing with time, and the transmission coefficient expression (5) is no longer applicable. In this case, the QEs in nanowire is described by semi-classical Dicke model, the Hamiltonian of the hybrid system is

\begin{equation} 
H = \frac{1}{2}\hbar\omega_AS_z+\hbar\Omega_R(t)(S_{\dagger}+S_-)\cos{(\omega t+\varphi)}
\label{Eq_6}
\end{equation}
Where\; $\Omega_R(t)=-\pmb{\mu}\cdot\pmb{F}(\pmb{r})|a(t)|/(\hbar\sqrt{\varepsilon_0})=\chi|a|$, \; $\pmb{\mu}$ is the transition dipole moment of a single QE (take $8\times 10^{-29} C\cdot m$ in this article), $\pmb{F}(\pmb{r})$ is the normalized distribution function of the electric field, $S_{\dagger}(t)=\sum_{j=1}^N S_{\dagger}^{(j)}(t)$ is the collective dipole operator. In the case of rotating wave approximation, the Hamiltonian in interaction picture at resonance ($\omega-\omega_A = 0$) can be expressed as $H_I=\hbar\Omega_R(t)(S_{\dagger}e^{-i\varphi}+S_-e^{i\varphi})/2$. It can be proved that $H_I(t_1)$ and $H_I(t_2)$ are commutative, that is $[H_I(t_1), H_I(t_2)]=0$. And the collective quantum operators with different subscripts ($i \neq j$) are also commutative, that is $[S_{\pm}^{(i)},S_{\pm}^{(j)}]=0$. According to the Glauber formula \cite{r39}, the time evolution operator in interaction picture is
\begin{equation}
U_I(t) = \prod \limits_{j=1}^N \exp{\Big\{-\frac{i}{2}\Big[e^{-i\varphi}S_{\dagger}^{(j)}+e^{i\varphi}S_-^{(j)}\Big]\int_0^t\Omega_R(t)dt\Big\}}
\label{Eq_7}
\end{equation}
We assume that the initial state of the QEs system is $\lvert\psi_0\rangle=\lvert g_1\rangle\lvert g_2\rangle\cdots\lvert g_N\rangle$, then the wave function of the QEs system is
\begin{equation}  
\lvert\psi\rangle = \prod\limits_{j=1}^N [\cos{(\frac{1}{2}\int_0^t \Omega_R dt)}\lvert g_j\rangle - i\sin{(\frac{1}{2}\int_0^t\Omega_Rdt)}e^{-i\varphi}\lvert e_j\rangle]
\label{Eq_8}
\end{equation}
Here, $\lvert g_j\rangle$ and $\lvert e_j\rangle$ are the wave functions of the ground state and excited state of the j-th QE, respectively. If the mode volume of the TMv mode in VG cavity is $V_{eff}$, the equivalent number density of the QEs in nanowire is $n = N/V_{eff}$. The energy of the QEs system is $\langle H \rangle=\langle\psi\lvert H \lvert\psi\rangle$, it is a function of time. Then, the output power density of the QEs system can be expressed as follows
\begin{equation}    
p_{QE}=-\frac{d\langle H \rangle}{V_{eff}dt} =-n\hbar\omega_A (\Omega_R/2)\sin\big(\int_0^t\Omega_R  dt\big)
\label{Eq_9}
\end{equation}
Equation (9) shows that the coupling between QEs and the VG cavity will make the QEs system to absorb or release energy, this energy will be converted into the electromagnetic field energy in VG cavity. This is equivalent to adding an additional field amplitude $a_{QE}$. So, the total field in the VG cavity is $a=a_1+a_{QE}$. Meanwhile, the electromagnetic field energy in VG cavity is also leaking out. The relationship between them is implied in the power conservation equation (13). Since the output power density of the QEs system induces an additional field $a_{QE}$, it is necessary to make the following appropriate corrections to the coupled mode theory (CMT) equation
\begin{align} 
\begin{split}
\frac{da_1}{dt}=-&(i\omega_1+\frac{1}{\tau_{10}}+\frac{1}{\tau_{11}}+\frac{1}{\tau_{12}})a_1+\kappa_{11}b_{1+}+\kappa_{12}b_{2+}
\end{split} \\ 
  b_{1-}&=-b_{1+}+\kappa_{11}^*(a_1+a_{QE})=-b_{1+}+\kappa_{11}^*a \\ 
  b_{2-}&=-b_{2+}+\kappa_{12}^*(a_1+a_{QE})=-b_{2+}+\kappa_{12}^*a
\end{align}
Where $a_1$ is the electric field of the TMv mode without nanowire in VG cavity, and $a$ is the total electric field in the resonator. For symmetrical systems, in the case of the rate of energy change in VG cavity is much smaller than the output rate and internal loss rate of the cavity, when the excitation light is resonated with the QEs and only injected from waveguide S$_1$ (that means $b_{2+}=0$), the power conservation is expressed as
\begin{equation}  
\lvert b_{1+} \rvert ^2 + p_{QE} = \lvert b_{1-} \rvert ^2 + \lvert b_{2-} \rvert ^2 + \frac{2}{\tau_{10}}\lvert a \rvert ^2 + \frac{d\lvert a \rvert ^2}{dt}
\end{equation}
Then, the following equation on the electric field amplitude can be obtained
\begin{equation} 
\frac{d\lvert a \rvert}{dt}+(\frac{2}{\tau_1}+\frac{1}{\tau_{10}})(|a|-|a_1|)+\frac{\chi}{4}\sin\Big(\int_0^t\Omega_R dt\Big) n\hbar\omega_A=0
\label{Eq_14}
\end{equation}
At this time, the transmittance of the hybrid system can be expressed as
\begin{equation} 
T=\Bigg\lvert \frac{b_{2-}}{b_{1+}}\Bigg\rvert ^2 = T_0\Bigg|\frac{a(t)}{a_1}\Bigg|^2
\label{Eq_15}
\end{equation}
$T_0$ is the transmittance of the system without QEs in nanowire.

Equation (15) shows that the transmittance of the hybrid system can be obtained as a function of time after solving the changing field amplitude $|a(t)|$ from equation (14). Moreover, the relationship between the field amplitude $|a_1|$ and the incident light intensity is $I_{in} = c\lvert a_1 \rvert ^2/(n_{eff}T_0)$, which indicates that the transmittance of the hybrid system is also related to the incident light intensity, $c$ is the speed of light in free space, $n_{eff}$ is the equivalent refractive index of the waveguide. Based on Eqs. (14) and (15), next investigated will be the influences of the incident light intensity and the QEs number density on the transmittance function of the hybrid system in the case of continuous light input.

Since the transmittance of the hybrid system is oscillating with time, two important parameters to describe the oscillation will be studied — pulse repetition frequency (PRF) and extinction ratio [$EXT = 10\log_{10}(T_{max}/T_{min}) = 10\log_{10}(a_{max}/a_{min})$]. The pseudo-color diagrams of the dependence of PRF and EXT on the incident light intensity and the QEs number density are shown in Figs. 3(a) and 3(b). Firstly, the point B (170, 1.8) is taken as an example to elucidate the formation mechanism of the femtosecond pulses with high repetition frequency.

\begin{figure}[bhtp]
\centering\includegraphics[width=8.7cm]{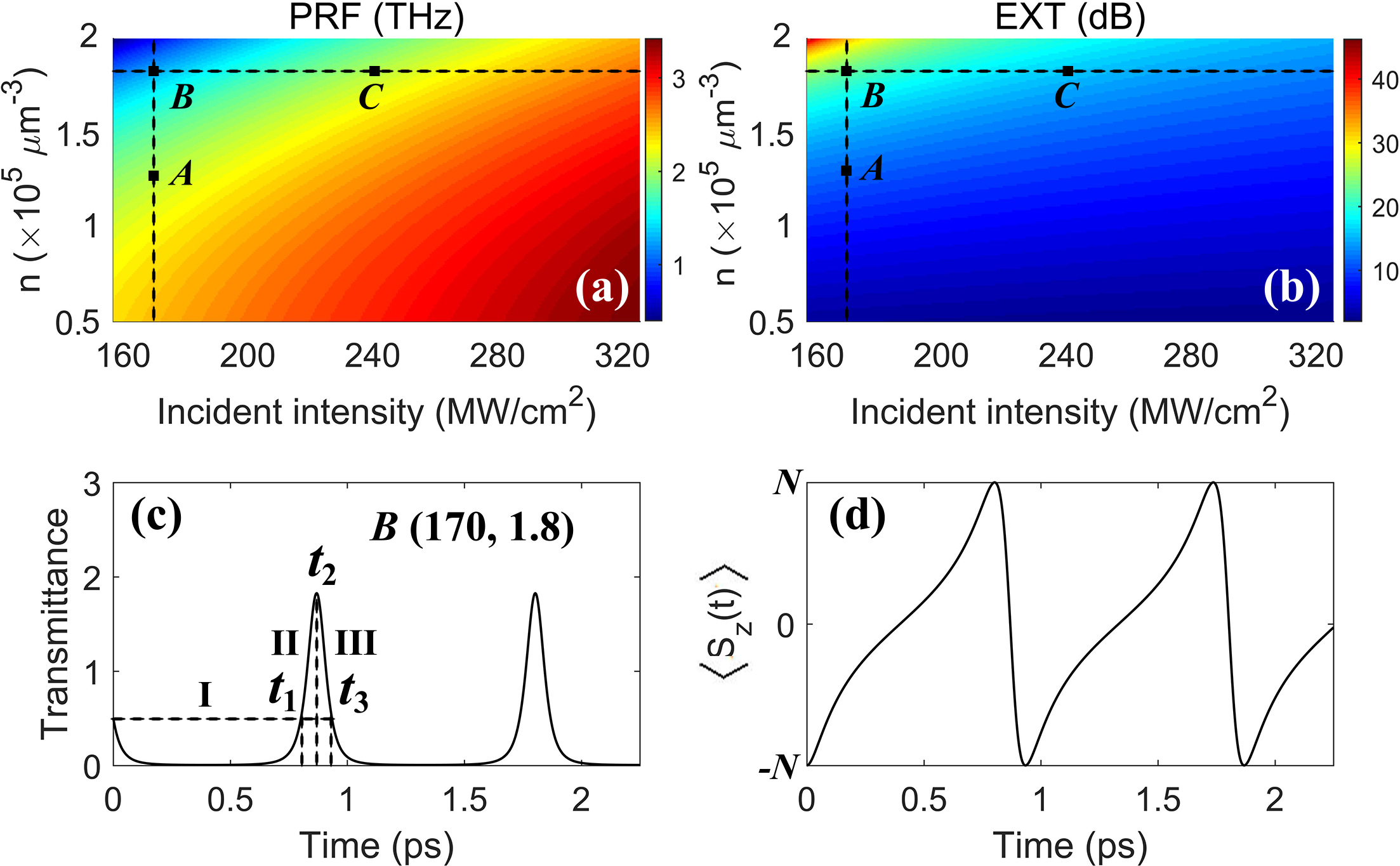}
\caption{(a) and (b) Pseudo-color diagrams of the dependence of PRF and EXT on the incident light intensity and the QEs number density. (c) and (d) Time evolution curves of the transmittance and the collective population difference function $\langle Sz(t) \rangle$ at point B. $\langle Sz(t) \rangle=\langle\psi\lvert \sum_{j=1}^N S_z^{(j)} \lvert\psi\rangle$, $S_z^{(j)}=\lvert e_j \rangle\langle e_j \lvert-\lvert g_j \rangle\langle g_j \lvert$, the wave function $\lvert\psi\rangle$ of the QEs system is given by Eq. (8). The equivalent refractive index of the waveguide is $n_{eff}=1.32$, initial transmittance is $T_0=0.5$. The nanowire is placed at the position of $\pmb{F}(\pmb{r})=1$.}
\label{Fig_3}
\end{figure}

The curve of the transmittance changing with time under the condition of point B is shown in Fig. 3(c). According to the expression (9) of the output power density, the phase function is defined as $\varphi(t)=\int_0^t\Omega_R dt$. According to the phase function, a period can be divided into three stages. The duration from 0 to $t_1$ is called stage I. In stage I, $0 < \varphi < \pi$, then $p_{QE} < 0$. This indicates that the QEs system is absorbing energy, so the additional field amplitude $a_{QE} < 0$, total field amplitude $a = a_1 + a_{QE} < a_1$, and it is known that $T < T_0$ from the transmittance expression (15). The duration from $t_1$ to $t_2$ is called stage II. In stage II, $\pi < \varphi < 3\pi/2$, $p_{QE} > 0$. This indicates that the QEs system is releasing energy, the additional field amplitude $a_{QE} > 0$, total field amplitude $a > a_1$, the transmittance of the system $T > T_0$. The duration from $t_2$ to $t_3$ is called stage III. Stage III is approximately the time mirror of stage II.

From 0 to $t_3$ is a pulse period. At the beginning, all QEs are in the ground state, due to the interaction between QEs and VG cavity, the QEs will absorb energy. Until the time of $t_1$, all QEs have transitioned to the excited state, the process of the QEs system absorbing energy is over. After that, also attribute to the interaction between QEs and VG cavity, in stages II and III the QEs releases the energy absorbed in stage I. The field amplitude in the plasmonic resonator will be caused to oscillate with time by such a process of absorbing and releasing energy. Moreover, during the whole process of stage I, the total field amplitude in plasmonic resonator is smaller than that in stages II and III. And it can be known combined with the integral expression of the phase function that the duration of stage I is naturally longer than the total durations of stages II and III, as shown in Fig. 3(c). The collective population difference function can also be affected by the oscillating field amplitude, of which the time evolution curve is given by Fig. 3(d). Since the total field amplitude in stage I is relatively small, the rate of QEs transition from the ground state to the excited state is relatively slow, so the duration is relatively long. In stages II and III, the total field amplitude in plasmonic resonator is larger, the rate of QEs transition from the excited state to the ground state is faster, thereby the duration is shorter. This results in the transmittance of the system to behave as the form of pulse oscillation with time. The PRF and EXT of the pulse in Fig. 3(c) are 1.07 THz and 26 dB respectively, and the pulse width is 88 fs, which shows that it is feasible to obtain femtosecond pulses with THz repetition frequency in transmission plasmonic systems.

\begin{figure}[htbp]
\centering\includegraphics[width=8.4cm]{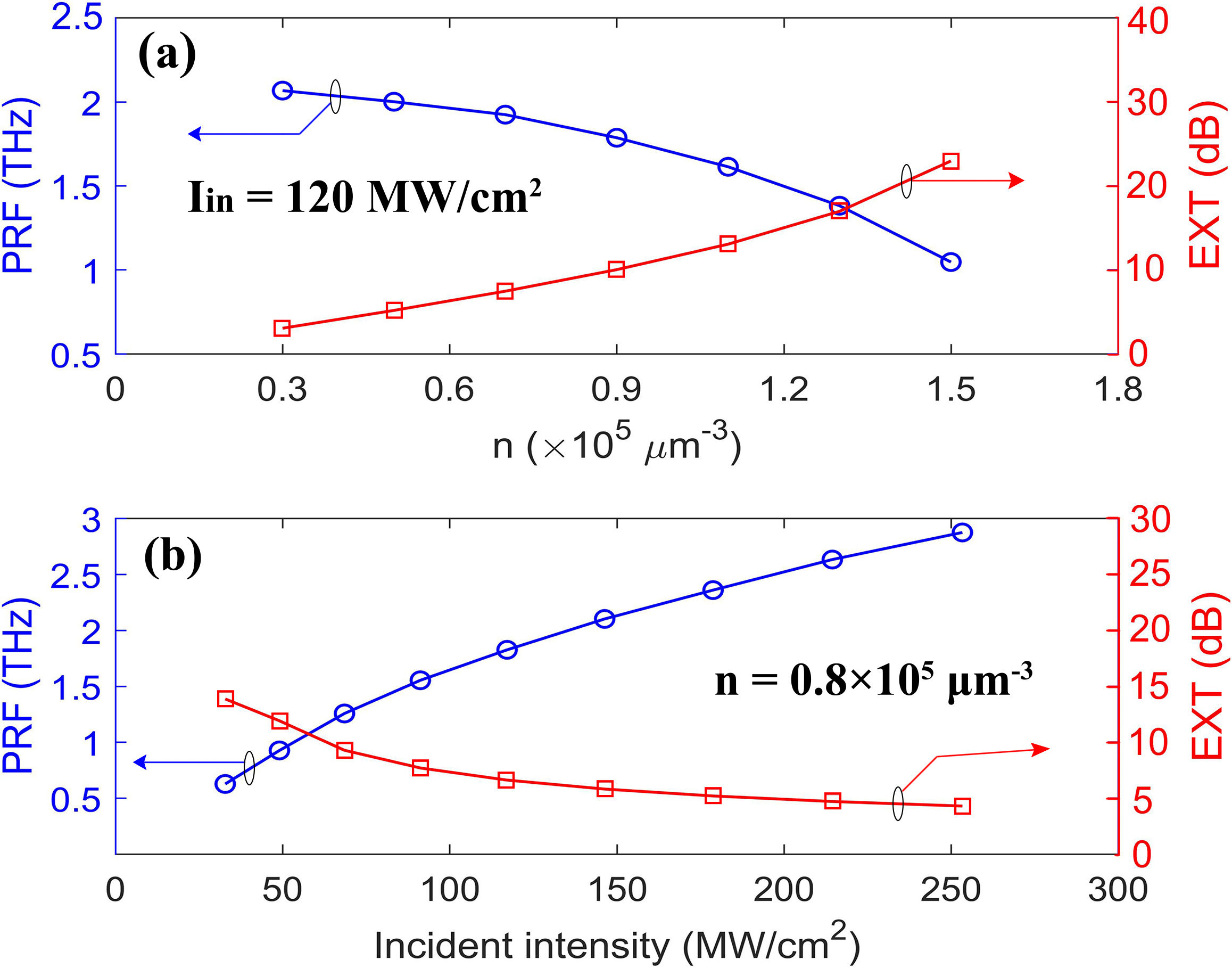}
\caption{(a) Curves of the PRF and EXT changing with the QEs number density when the incident light intensity is 120 MW/cm$^2$, respectively. (b) Curves of the PRF and EXT changing with the incident light intensity when the QEs number density is 0.8 $\times 10^5$ $\mu$m$^{-3}$, respectively.}
\label{Fig_4}
\end{figure}

The pseudo-color images of Figs. 3(a) and 3(b) show that the PRF and EXT of the pulse oscillation are closely related to the incident light intensity and the QEs number density. Next, the influence of the incident light intensity and the QEs number density on the PRF and EXT of the hybrid system will be studied respectively.

First studied is the influence of the QEs number density on PRF and EXT when the incident light intensity is constant. As shown in Fig. 4(a), when the incident light intensity is 120 MW/cm${^2}$, the larger the QEs number density, the smaller the PRF and the lager the EXT. When the incident light intensity is constant, the larger the QEs number density, the greater the absorbed power of the QEs system in stage I, resulting in smaller total field amplitude in the resonator, and the duration of stage I will be longer. However, the time changes of stage II and III are much smaller than that of stage I. Overall, the time of a pulse period will become longer and the PRF will decrease. For extinction ratio, the larger the QEs number density, the smaller the minimum value of the total field amplitude in stage I, and the greater the output power density of the QEs system in stages II and III which results in greater total field amplitude at $t_2$, so the EXT will increase. When the QEs number density is close to 0, the output power of the QEs system will also approach 0, the field amplitude in the resonator tends to the constant $|a_1|$, so the EXT will approach zero and the PRF tends to a limit value $\chi|a_1|/(2\pi)$.

Next analyzed is the influence of the incident light intensity on PRF and EXT when the QEs number density is fixed. As shown in Fig. 4(b), when the QEs number density is $n = 0.8 \times 10^{-5} \mu m^{-3}$, the PRF will increase but the EXT will decrease with the increase of the incident light intensity. When the QEs number density is constant, the stronger the incident light intensity, the greater the total field amplitude in all three stages, and the shorter the time for the phase to change from 0 to $2\pi$, and thus the PRF will increase. From Eqs. (9) and (12), it can be known that the ratio of the output power of the QEs system to the transmission power is inversely proportional to the total field amplitude in VG cavity, that is $p_{QE}/|b_{2-}|^2 \propto 1/|a(t)|$. So, the larger the total field amplitude $|a(t)|$, the smaller the ratio  $p_{QE} / |b_{2-}|^2$, which indicates that the stronger the incident light intensity, the weaker the influence of the QEs system on the transmission response, and the smaller the peak-to-valley ratio of the transmittance curve, naturally the EXT will decrease. When the incident light intensity is extremely strong, the influence of the QEs system on the transmission response of the hybrid system can be ignored, the transmittance of the system becomes the constant $T_0$, and the PRF and EXT respectively tend to $\chi|a_1|/(2\pi)$ and 0.

The semi-classical Dicke model is no longer applicable when the ratio of the QEs number density to the incident light intensity is extremely large  (this means $n\rightarrow\infty$ or $I_{in}\rightarrow 0$), the full quantum theory which is more accurate is required for analyzing. Moreover, the plasmon mode volume will become lager when the QEs are relaxed in all 3D directions, and resulting in lower coupling efficiency. Therefore, the plasmon mode should be limited in more dimensions to reduce the mode volume and improve the coupling efficiency.

\section{Conclusion}
In summary, we studied the transmission response of a hybrid system consisting of plasmonic resonator and a nanowire embedded with two-level QEs. Investigations under the weak excitation limit show that the coupling between QEs and the TMv mode of the VG cavity can reach the region of strong coupling. In the case of continuous wave incidence, attribute to the presence of QEs, the transmittance of the hybrid system behave as the form of pulse oscillation with time. The pulse repetition frequency could reach the magnitude of terahertz, simultaneously, the pulse width is below 100 fs and the extinction ratio also reaches very high values. Furthermore, to be different from the method of modulating plasmonic femtosecond pulse in the reports of Sámson \cite{r40} and Kim \cite{r41} \textit{et al.}, we can freely control the pulse repetition frequency and extinction ratio by the incident light intensity and QEs number density to obtain the pulse oscillation required. This provides a feasible route for achieving ultra-short pulses with high repetition frequency at the nanoscale, which also has potential applications in optical clock signals.

\hspace*{\fill} \\
\noindent \textbf{Funding.} National Key Research and Development Program of China (2016YFA0301300); Fundamental Research Funds for the Central Universities.

\hspace*{\fill} \\
\noindent \textbf{Disclosures.} The authors declare no conflicts of interest.

\end{document}